# A New Transferable Interatomic Potential for Molecular Dynamics Simulations of Borosilicate Glasses

Mengyi Wang[1], N. M. Anoop Krishnan[1,2], Bu Wang[1], Morten M. Smedskjaer[3], John C. Mauro[4], Mathieu Bauchy[1]

[1] Physics of AmoRphous and Inorganic Solids Laboratory (PARISlab), Department of Civil and Environmental Engineering, University of California, Los Angeles, California, 90095, USA
[2] Department of Civil Engineering, Indian Institute of Technology Delhi, Hauz Khas, New Delhi 110016, India
[3] Department of Chemistry and Bioscience, Aalborg University, 9220 Aalborg, Denmark
[4] Department of Materials Science and Engineering, The Pennsylvania State University, University Park, Pennsylvania 16802, USA

## Abstract

Borosilicate glasses are traditionally challenging to model using atomic scale simulations due to the composition and thermal history dependence of the coordination state of B atoms. Here, we report a new empirical interatomic potential that shows a good transferability over a wide range of borosilicate glasses—ranging from pure silicate to pure borate end members—while relying on a simple formulation and a constant set of energy parameters. In particular, we show that our new potential accurately predicts the compositional dependence of the average coordination number of boron atoms, glass density, overall short-range and medium-range order structure, and shear viscosity values for several borosilicate glasses and liquids. This suggests that our new potential could be used to gain new insights into the structure of a variety of advanced borosilicate glasses to help elucidate composition-structure-property relationships—including in complex nuclear waste immobilization glasses.

## 1. Introduction

Borosilicate glasses have various applications, including kitchen and laboratory glassware [1,2], glass fibers [3], chemically strengthened protective cover glasses [4–6], glass substrates for high performance displays [7], and nuclear waste immobilization [8–10]. Their atomic structure is partially derived from those of pure glassy $SiO_2$ and $B_2O_3$ [11]. The atomic network of $SiO_2$ consists of $SiO_4$ tetrahedra that are interconnected through their corners by bridging oxygen (BO) atoms. The structure of pure $B_2O_3$ glass is composed of triangular $BO_3$ structural units connected by BOs at their corners. Starting from these basic topologies, the introduction of network modifiers— such as alkali or alkaline-earth cations—can have different effects. Specifically, each alkali cation M may associate with either (i) Si or B to create a non-bridging oxygen (NBO) atom or (ii) with B to convert boron from a trigonal $BO_3$ unit to a $BO_4$ tetrahedral unit by acting as a charge compensator. While the latter mechanism is typically predominant at M/B < 0.5, both mechanisms can coexist [12]. The effect of alkaline-earth cations R is typically equivalent to that of two alkali cations, i.e., they can form two NBOs, stabilize two $BO_4$ units, or one of each [13].

Developing novel modified borosilicate glasses with improved properties (e.g., with enhanced damage resistance, suppressed relaxation behavior, or enhanced resistance to corrosion in the context of nuclear waste immobilization) requires an accurate knowledge of their atomic structure. To this end, classical molecular dynamics (MD) simulations have been extensively used to explore the structure and properties of various silicate glasses with great success [14–17]. However, classical MD simulations of borosilicate glasses are notoriously challenging due to the composition and thermal history dependence of the coordination state of B atoms. As such, previous simulations of borosilicate glasses relying on classical empirical interatomic potentials resulted in only a partial agreement with experimental data [18–20]. Although *ab initio* simulations can overcome the limitations of classical potentials and properly describe the coordination number of B cations in borosilicate liquids, they are restricted to shorter timescales than classical MD simulations and, hence, significantly higher cooling rates [21]. Hence, they tend to yield glasses exhibiting very high fictive temperature, which, in turn, affects the average coordination number of the B cations [21,22].

To overcome the limitations of classical potentials while retaining their computational efficiency, Kieu *et al.* recently introduced a new empirical interatomic potential (referred to as Kieu's potential hereafter) allowing them to reproduce the variations in the coordination number of B cations as a function of composition [23]. This potential was later extended to include Al atoms [24,25]. The main idea of this potential is to rely on non-constant partial charges and energy parameters for B cations. These parameters are empirically adjusted based on the glass composition to match experimental results. Although Kieu's potential offers an excellent agreement with experimental data for the glass compositions it is fitted against, it comes with several issues as follows. (i) Adjusting the partial charge of B cations requires recalibrating the charges of all the other elements (Si, O, Na, etc.) in order to maintain the charge neutrality of the system. As such, variations in the average coordination number of B cations affect the interatomic interactions between all the other pairs of atoms, which has an unclear physical meaning. (ii) Kieu's potential is heavily parametrized (for instance, 7 parameters are required to calculate the charge of B cations), thereby making the calibration for new systems tedious. (iii) The potential lacks an intrinsic predictive power with respect to composition–structure relationships, since the coordination of B cations is parametrized from experimental data. (iv) Finally, as shown below, it is poorly transferable to new compositions that were not initially included during its parametrization.

Here, we follow an alternative route and present a new empirical interatomic potential that exhibits a good transferability to a wide variety of borosilicate glasses while retaining a simple formulation and constant parameters. We show that our new potential is able to properly predict the compositional dependence of several structural and dynamical properties—such as density, average coordination number of B cations, pair distribution functions, and viscosity—for a large selection of borosilicate compositions ranging from borate to silicate glasses.

## 2. Simulation methodology
### 2.1 Glass simulations
Our new potential is parametrized and validated based on the MD simulations of a series of borosilicate glasses, whose compositions and naming are presented in **Tab. 1**. This list comprises a series of borosilicate glasses with varying Si/B molar ratios at fixed network modifiers content (from 75B to 0B, ranging from modified silicates to modified borates). These glasses are selected as they have been extensively characterized experimentally by Smedskjaer and Mauro [26,27]. We also consider an additional glass (10B) as its structure was investigated by neutron diffraction experiments [13].

**Tab. 1:** Compositions of the glasses simulated herein.

| Glass ID | Chemical composition (mol %) | | | |
| --- | --- | --- | --- | --- |
| | $SiO_2$ | $B_2O_3$ | $Na_2O$ | $CaO$ |
| 75B | 0 | 75 | 15 | 10 |
| 62B | 13 | 62 | 15 | 10 |
| 50B | 25 | 50 | 15 | 10 |
| 37B | 38 | 37 | 15 | 10 |
| 24B | 51 | 24 | 15 | 10 |
| 12B | 63 | 12 | 15 | 10 |
| 6B | 69 | 6 | 15 | 10 |
| 0B | 75 | 0 | 15 | 10 |
| 10B | 60 | 10 | 15 | 15 |

These glasses are prepared and studied using MD simulations relying on the LAMMPS package [28]. The empirical potential used herein is described in **Sec.** 2.2. We use a cutoff of 11 Å both for the short-range and Coulombic interactions. The long-range Coulombic interactions are calculated with the PPPM algorithm with an accuracy of $10^{-5}$. The timestep is fixed at 1.0 fs. The glasses are simulated using the traditional melt-quench procedure as follows [29–31]. First, around 3000 atoms are randomly placed within a cubic box while ensuring the absence of any unrealistic overlap. The system is then melted at 3000 K in the canonical (*NVT*) ensemble for 10 ps and at zero pressure (*NPT* ensemble) for 100 ps, which ensures a complete loss of the memory of the initial configuration. The system is subsequently cooled linearly to 300 K at zero pressure (*NPT* ensemble) with a cooling rate of 1 K/ps. All of the resulting glasses are further relaxed at 300 K and zero pressure for 100 ps before an *NVT* run of 100 ps for statistical averaging. In the following, all properties referring to the "glassy state" are averaged over 100 configurations extracted with an interval of 1 ps from this run. In specific cases, we observe that initial configurations with unrealistic structure tend to "explode" (i.e., their volume indefinitely increases over time) at high temperature in the *NPT* ensemble. In such cases, we first create a more realistic structure by melting the initial configuration at 3000 K in the *NVT* ensemble for 100 ps and cooling the system linearly to 300 K at fixed volume (*NVT* ensemble). The obtained glassy structure is then used as the starting configuration for the melt-quench procedure previously described.

*2.2 Potential parametrization*
In the following, we detail the potential parametrization methodology that was followed. Our goal was to develop an empirical interatomic potential that: (i) exhibits a good transferability to a large selection of borosilicate glasses, (ii) relies on a constant set of parameters (i.e., with fixed partial charges and energy terms—in contrast with Kieu's potential [23]), and (iii) retains a simple two-body formulation and high computational efficiency. Indeed, as discussed in **Sec. 1**, although using composition-dependent parameters as in Kieu's potential can certainly allow one to improve the level of agreement with experiments for each individual composition [23], this approach intrinsically limits the transferability and the applicability of the potential to new and yet-unexplored compositions. Further, although relying on more elaborated potential formulations (e.g., three-body interactions [32], polarizable force fields [33], coordination-dependent charge-transfer potential [34], or reactive potentials [35–37]) could result in an improved description of the structure of oxide systems, these complexities come with additional computational cost. Here, we favor transferability and computational efficiency (potentially at the expense of fine accuracy for each specific system) by relying only on two-body Buckingham potential energy terms with fixed parameters:

$$U_{ij}(r_{ij}) = \frac{z_i z_j}{r_{ij}} + A_{ij}\exp\left(\frac{-r_{ij}}{\rho_{ij}}\right) - \frac{C_{ij}}{r_{ij}^6} \qquad \text{Eq. 1}$$

where $r_{ij}$ is the distance between the atoms *i* and *j*, $z_i$ the effective partial charge of atom *i*, and $A_{ij}$, $\rho_{ij}$, and $C_{ij}$ the energy parameters for the pair of atoms (*i*, *j*). The energy terms correspond to the Coulombic interactions, short-range electronic repulsion, and Van der Waals interactions, respectively. Note that, in general, an additional short-range repulsive term of the form $U(r) = B/r^n$ might be needed to avoid the "Buckingham catastrophe" at high temperature [31,38]. However, such a term is not needed here as the simulated systems exhibit a low glass transition temperature, so that an initial temperature as low as 3000 K is high enough to fully randomize the initial configuration within a few picoseconds.

To ensure a large transferability, our interatomic potential is largely based on the parametrization initially introduced by Guillot and Sator (referred to as the GS potential hereafter), as this potential has be proven to show an excellent transferability over a wide range of modified silicate minerals and glasses while retaining constant parameters [39–41]. Our potential is designed to be compatible with the GS potential to retain its proven transferability. Unfortunately, the original GS potential did not contain any interaction parameters for B cations. As such, we keep all the original parameters of the GS and restrict the addition of new terms to those involving B cations only (for instance, we do not add any Si–Si interaction terms, so that our new parametrization is fully equivalent to the GS potential in the absence of B cations). The partial charges used in the GS potential are also retained (see **Tab. 2**).

Specifically, our new parametrization differs from that of the GS potential through the addition of new energy terms for B–O, B–B, and B–Si energy terms. The B–Si terms are sourced from Kieu's potential. The B–O parameters are then adjusted to properly reproduce the evolution of the coordination number of B cations within the list of glasses presented in **Tab. 1**. The optimal B–O parameters are determined by minimizing the mean square difference between experimental and computed coordination numbers over the entire range of compositions. Finally, the B–B parameters are adjusted to properly reproduce the evolution of density within the list of glasses presented in **Tab. 1**. Again, the optimal B–B parameters are determined by minimizing the mean square difference between experimental and computed densities over the entire range of compositions. A few iterations of this procedure are performed until we obtain optimal values for the B–O and B–B terms. The full set of parameters is listed in **Tab. 3**—note that only electrostatic interactions are considered for all the pairs that are not listed (e.g., B–Na). An example LAMMPS input file is also provided in supplementary information.

**Tab. 2:** Fixed partial charge attributed to each element.

| Element | Partial charge (e) |
|---|---|
| O | −0.945 |
| Si | 1.89 |
| B | 1.4175 |
| Ca | 0.945 |
| *Na | 0.4725 |
| *Ti | 1.89 |
| *Al | 1.4175 |
| *$Fe^{3+}$ | 1.4175 |
| *$Fe^{2+}$ | 0.945 |
| *Mg | 0.945 |
| *K | 0.4725 |

* These parameters are sourced from the original Guillot–Sator interatomic potential [39]. They are indicated herein for reference, although these elements are not considered in the present study.

**Tab. 3:** Parameters of the interatomic potential.

| Bond | $A_{ij}$ (eV) | $\rho_{ij}$ (Å) | $C_{ij}$ (eV·Å$^6$) |
|---|---|---|---|
| O–O | 9022.79 | 0.265 | 85.0921 |
| Si–O | 50306.10 | 0.161 | 46.2978 |
| B–O | 206941.81 | 0.124 | 35.0018 |
| B–B | 484.40 | 0.35 | 0.0 |
| Si–B | 337.70 | 0.29 | 0.0 |
| Na–O | 120303.80 | 0.17 | 0.0 |
| Ca–O | 155667.70 | 0.178 | 42.2597 |
| *Ti–O | 50126.64 | 0.178 | 46.2978 |
| *Al–O | 28538.42 | 0.172 | 34.5778 |
| *$Fe^{3+}$–O | 8020.27 | 0.19 | 0.0 |
| *$Fe^{2+}$–O | 13032.93 | 0.19 | 0.0 |
| *Mg–O | 32652.64 | 0.178 | 27.2810 |
| *K–O | 2284.77 | 0.29 | 0.0 |

* These parameters are sourced from the original Guillot–Sator interatomic potential [39]. They are indicated herein for reference, although these elements are not considered in the present study.

## 2.3 Computation of the structure and properties of glasses and liquids

### a. Pair distribution functions

To compare the structure of the simulated glasses with experimental data from neutron diffraction, we first compute each partial pair distribution functions (PDFs) $g_{ij}(r)$. The neutron PDF is then calculated as:

$$g_N(r) = (\sum_{i,j=1}^{n} c_i c_j b_i b_j)^{-1} \sum_{i,j=1}^{n} c_i c_j b_i b_j g_{ij}(r) \quad \text{Eq. 2}$$

where $c_i$ is the fraction of $i$ atoms ($i$ = O, Si, B, Ca, or Na) and $b_i$ the neutron scattering length of the species (given by 5.803, 4.1491, 5.30, 4.70, and 3.63 fm for O, Si, B, Ca, and Na cations, respectively). To enable a meaningful comparison with experimental data [42], the neutron PDF is subsequently broadened to account for the maximum of wave vector $Q_{max}$ used experimentally [42]. This is achieved by convoluting the computed neutron PDF with a normalized Gaussian distribution with a full width at half-maximum (FWHM) given by FWHM = $5.437/Q_{max}$ [31,42].

### b. Structure factor

To investigate the structure of the glasses over intermediate length scales, we compute the partial structure factors $S_{ij}(Q)$ from the Fourier transform of the partial PDFs:

$$S_{ij}(Q) = 1 + \rho_0 \int_0^R 4\pi r^2 (g_{ij}(r) - 1) \frac{\sin(Qr)}{Qr} F_L(r) dr, \quad \text{Eq. 3}$$

where $Q$ is the scattering vector, $\rho_0$ is the average atomic number density, and $R$ is half of the simulation box length. The $F_L(r) = \sin(\pi r/R)/(\pi r/R)$ term is a Lorch-type window function used to reduce the effect of the finite cutoff of $r$ in the integration [30]. The use of this function reduces the ripples at low $Q$ but can induce a broadening of the structure factor peaks. The total neutron structure factor is then evaluated from the partial structure factors following:

$$S_N(Q) = (\sum_{i,j=1}^{n} c_i c_j b_i b_j)^{-1} \sum_{i,j=1}^{n} c_i c_j b_i b_j S_{ij}(Q) \quad \text{Eq. 4}$$

### c. Coordination numbers

The connectivity of the network is assessed by computing the coordination numbers of Si and B cations. This is achieved by enumerating the number of O neighbors for each cation within its first coordination shell, with a cutoff chosen as the first minimum after the first peak of the partial PDF. This analysis is used to discern 3-fold from 4-fold coordinated B cations.

### d. Ring size distribution

The medium-range order structure of the glasses is further investigated by computing their ring size distributions [43]. All the ring size distribution computations are carried out using the RINGS package [44]. Here, rings are defined as the shortest closed paths within the borosilicate network, their size being defined as the number of Si or B atoms belonging to a given ring.

*e. Shear viscosity*

The ability of our potential to predict realistic dynamic properties is assessed by computing the shear viscosity of the borosilicate liquids at different temperatures by using the Green–Kubo formalism. This approach is based on the calculation of the autocorrelation function of the deviatoric stress tensor [40]:

$$F(t) = \langle P_{\alpha\beta}(t) P_{\alpha\beta}(0) \rangle \qquad \text{Eq. 5}$$

where the brackets refer to an average over the whole system. The $P_{\alpha\beta}$ terms are the off-diagonal components of the stress tensor ($\alpha, \beta = x, y, z$) defined as:

$$P_{\alpha\beta} = \sum_{i=1}^{N} m_i v_i^\alpha v_i^\beta + \sum_{i=1}^{N} \sum_{j>i}^{N} F_{ij}^\alpha r_{ij}^\beta \text{ with } \alpha \neq \beta \qquad \text{Eq. 6}$$

where $N$ is the number of atoms, $m_i$ is the mass of atom $i$, $v_i^\alpha$ is the $\alpha$ component of the velocity of atom $i$, $F_{ij}^\alpha$ is the $\alpha$ component of the force applied by atom $i$ on atom $j$, and $r_{ij}^\alpha$ is the $\alpha$ component of distance vector from atom $i$ to atom $j$. The shear viscosity is then calculated from the time integral of the autocorrelation function of the deviatoric stress tensor [40]:

$$\eta = \frac{1}{k_B T V} \int_0^\infty F(t) \mathrm{d}t \qquad \text{Eq. 7}$$

where $k_B$ is the Boltzmann constant, $T$ the temperature, and $V$ the volume.

## 3. Results

*3.1 Parametrization of the potential*

We first assess the success of the potential parametrization. As detailed in **Sec. 2.2**, our new potential was parametrized based on the experimental density and average B coordination number [26] of a series of borosilicate glasses (see **Tab. 1**). As shown in **Fig. 1**, our new potential offers an excellent agreement with these experimental data.

As shown in **Fig. 1a**, we observe that the glass density gradually increases with the amount of $SiO_2$, reaches a maximum at around $[SiO_2]/([SiO_2]+[B_2O_3]) = 0.7$, and eventually decreases at high amount of $SiO_2$. We note that the computed density values are in good agreement with experimental data, both in terms of absolute values and trend. We observe a slight discrepancy between computed and experimental density values at high amount of $SiO_2$, that is, at low $B_2O_3$ concentration. This arises from the fact that, in this domain, our new parametrization is essentially irrelevant as the newly added energy terms only involve B atoms. As such, the density values obtained at high amount of $SiO_2$ are controlled by the original GS potential.

The fraction of 4-fold coordinated B atoms is found to increase monotonically with [SiO$_2$] (see **Fig. 1b**). This can be understood from the fact that, at high amount of B$_2$O$_3$ (i.e., low amount of SiO$_2$), there is a deficit of Na and Ca cations as compared to the large number of B atoms, so that only a small fraction of the B atoms can be charge-compensated and form BO$_4$ units. On the other hand, at low amount of B$_2$O$_3$ (i.e., high amount of SiO$_2$), there is a large excess of Na and Ca atoms, so that the majority of the B atoms tend to form charge-compensated BO$_4$ units. Again, we observe an excellent agreement with experimental values, both in terms of absolute values and trend.

Although the agreement between experimental and computed density and coordination data is not surprising as these data are explicitly used to parametrize the potential, it is still remarkable that our new potential is able to properly describe the behavior of both these quantities over such an extended computational range (from borate to silicate end-member glasses) while relying only on constant charges and potential parameters.

For comparison, **Fig. 2** shows the same experimental data compared with the computed data obtained by using Kieu's potential [23]. Note that the original Kieu's potential does not contain any Ca–O interactions terms [23]. To this end, here we slightly modified the equation used to determine the partial charge and energy terms of B atoms with the assumption that each Ca cation has the same effect as that of two Na cations. As shown in **Fig. 2**, Kieu's potential does not succeed in properly predicting the evolution of the density and coordination values. Specifically, we observe a non-monotonic evolution of the density and coordination values, which may arise from the discontinuous form of the equation used to determine the energy terms as a function of composition [23]. This illustrates the fact that, although Kieu's potential offers an excellent agreement with experimental data for the compositions that were explicitly used for its parametrization, it might not yield realistic results when extrapolated to new unexplored compositions.

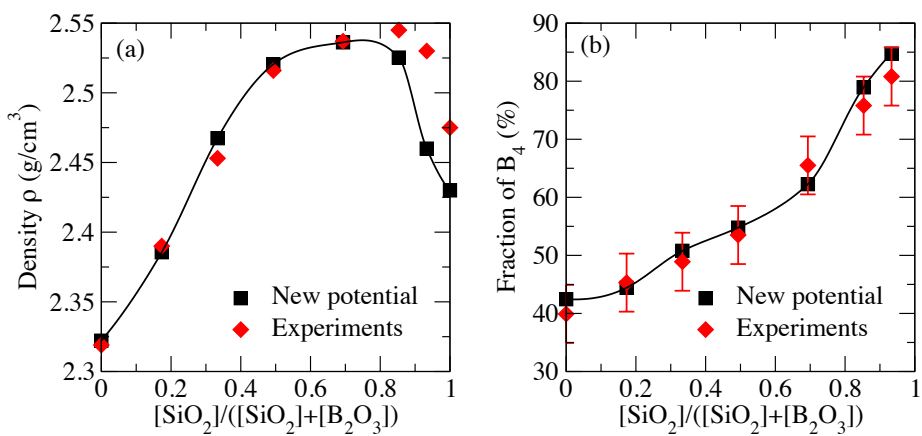

**Fig. 1: (a)** Density and **(b)** fraction of 4-fold coordinated B cations computed with our new potential and compared with experimental data [26]. Error bars on the density values are smaller than the size of the symbols. The line is a guide for the eye.

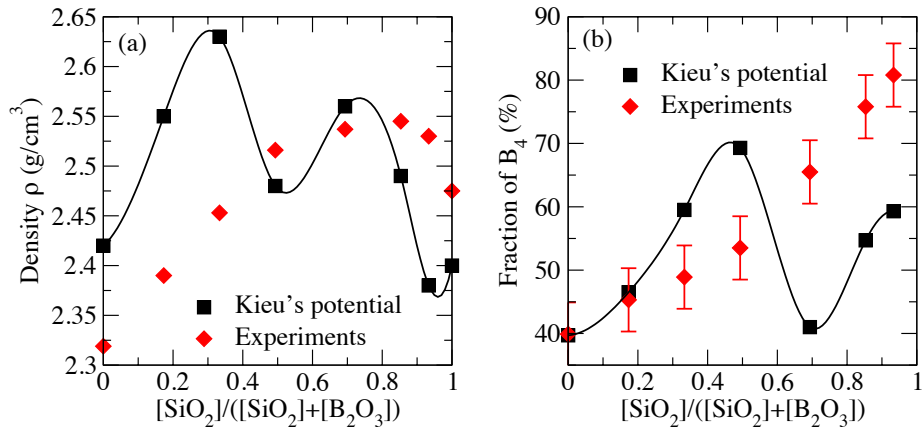

**Fig. 2: (a)** Density and **(b)** fraction of 4-fold coordinated B cations computed with Kieu's potential and compared with experimental data [26]. The line is a guide for the eye.

## *3.2 Pair distribution functions*

We now assess the ability of our new potential to predict a realistic overall structure for borosilicate glasses. To this end, we compare the outcomes of our potential to neutron diffraction data [13] obtained in the case of a sodium-calcium borosilicate glass (glass B10, see **Tab. 1**). Except for the coordination data presented in **Sec. 3.1**, no other structural data were included in the parametrization of the potential, so that the comparison with neutral diffraction data constitutes a real test for our new potential. In addition, note that the B10 glass is not part of the compositional range used for the training of our potential so that the comparison with the neutron diffraction data also allows us to assess the performance of our potential when extrapolated to a new composition.

**Fig. 3** shows the computed neutron PDF (see **Sec. 2.3**) of the 10B glass, compared with experimental data. Overall, we observe an excellent agreement between computed and experimental data, which suggests that our new potential offers a realistic description of the short-range order structure of borosilicate glasses. Specifically, we observe that the position and intensity of the correlation peaks are well predicted. Note that a shoulder located around *r* = 2.2 Å in experiments is not well reproduced. This appears to be a general limitation of classical potential as a similar discrepancy is observed in sodium silicate glasses [30,31,45]. One cannot also exclude that it might be spurious ripple introduced by the Fourier transform performed during the analysis of the experimental results.

The structure predicted by Kieu's potential is also presented in **Fig. 3** for comparison. Overall, we observe a reasonable agreement between simulation and experimental results. However, we note that the positions of the peaks predicted by Kieu's potential are slightly underestimated. Specifically, since the first peak around 1.6 Å mostly arises from Si–O correlations, this discrepancy suggests that Kieu's potential does not properly describe the environment of Si atoms. This might arise from the fact that, in Kieu's approach, the addition of B atoms effectively results in a modification of the charges of Si and O atoms to maintain the electroneutrality of the system. The consequence of this variation in the charges of Si and O atoms is unclear as fixed charges were used during the parametrization of the original GS potential.

The comparison between experimental and computed results yields a $R_\chi$ factor [29,36,42,45] of 6.1 and 14 % in the case of our new potential and Kieu's potential, respectively. Note that a value of around 10 % is typically considered satisfactory for classical MD simulations [42]. This demonstrates the superiority of our new potential in predicting a realistic short-range order structure for this glass. Once again, this illustrates the fact that, although Kieu's potential offers an excellent description of the structure of the glasses used during its parametrization [23], its transferability to new compositions is limited.

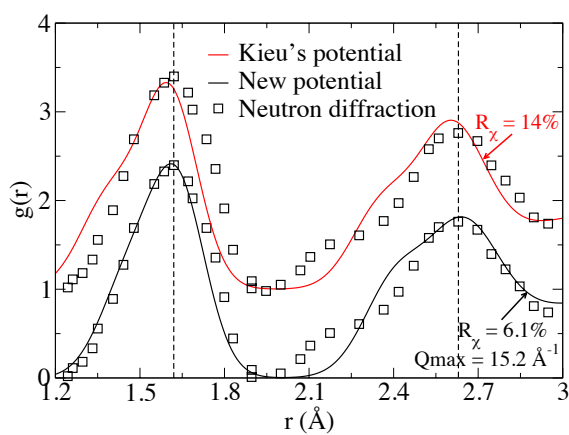

**Fig. 3:** Neutron pair distribution function of the 10B glass computed with our new potential and compared with neutron diffraction data [13]. The total neutron pair distribution function of the same glass computed with Kieu's potential [23] is also presented and compared with the same experimental data (vertically shifted by +1).

We now investigate in more detail the environment of each element by computing the partial PDFs (see **Sec. 2.3**). **Fig. 4** shows the partial Si–O, B–O, Ca–O, Na–O, Ca–Ca, and Na–Na PDFs. As shown in **Fig. 4a**, the local environment of Si atoms is largely unaffected by composition, although some variations are observed within the second coordination shell. The average Si–O bond distance is consistently found at around 1.63 Å, in good agreement with experimental values [13]. In contrast, the first peak of the B–O PDF shows a bimodal distribution and the average B–O bond distance tends to decrease upon the increase in the amount of $B_2O_3$ (see **Fig. 4b**). This arises from the gradual increase in the population of 3-fold coordinated B atoms at the expense of 4-fold coordinated species as the amount of $B_2O_3$ increases (see **Fig. 1b**). We observe that the average B–O bond length is around 1.40 and 1.46 Å for 3- and 4-fold coordinated B atoms, respectively, in good agreement with experimental values [13].

As shown in **Fig. 4c-d**, the local environment of the network modifiers appears to be more sensitive to the glass composition. In both cases, we observe that the interatomic bond distance significantly increases upon the addition of $B_2O_3$. This suggests that the Ca–O and Na–O bond distances are larger when Ca and Na act as charge-compensators (i.e., to compensate the charge of negatively-charged $BO_4$ tetrahedral units) than when they act as network-modifiers (i.e., to create NBOs). This can be understood from the fact that Ca and Na cations are more strongly bonded to the NBO(s) they create [46] than to the surrounding BOs when they act as charge-compensators.

Finally, we observe the existence of some spatial correlations among Ca and Na cations, which manifests itself as an intense correlation peak around 3.5 and 3.1 Å for Ca and Na cations, respectively (see **Fig. 4e-f**). This suggests that Ca and Na cations tend to cluster within pockets or channels [31,47]. However, the propensity for cations agglomeration appears to decrease upon the addition of $B_2O_3$, as the intensity of the correlation peak tends to decrease. Note that the details of the spatial distribution of the modifiers in borosilicate glasses is beyond the scope of the present manuscript and will be further investigated in a separate study.

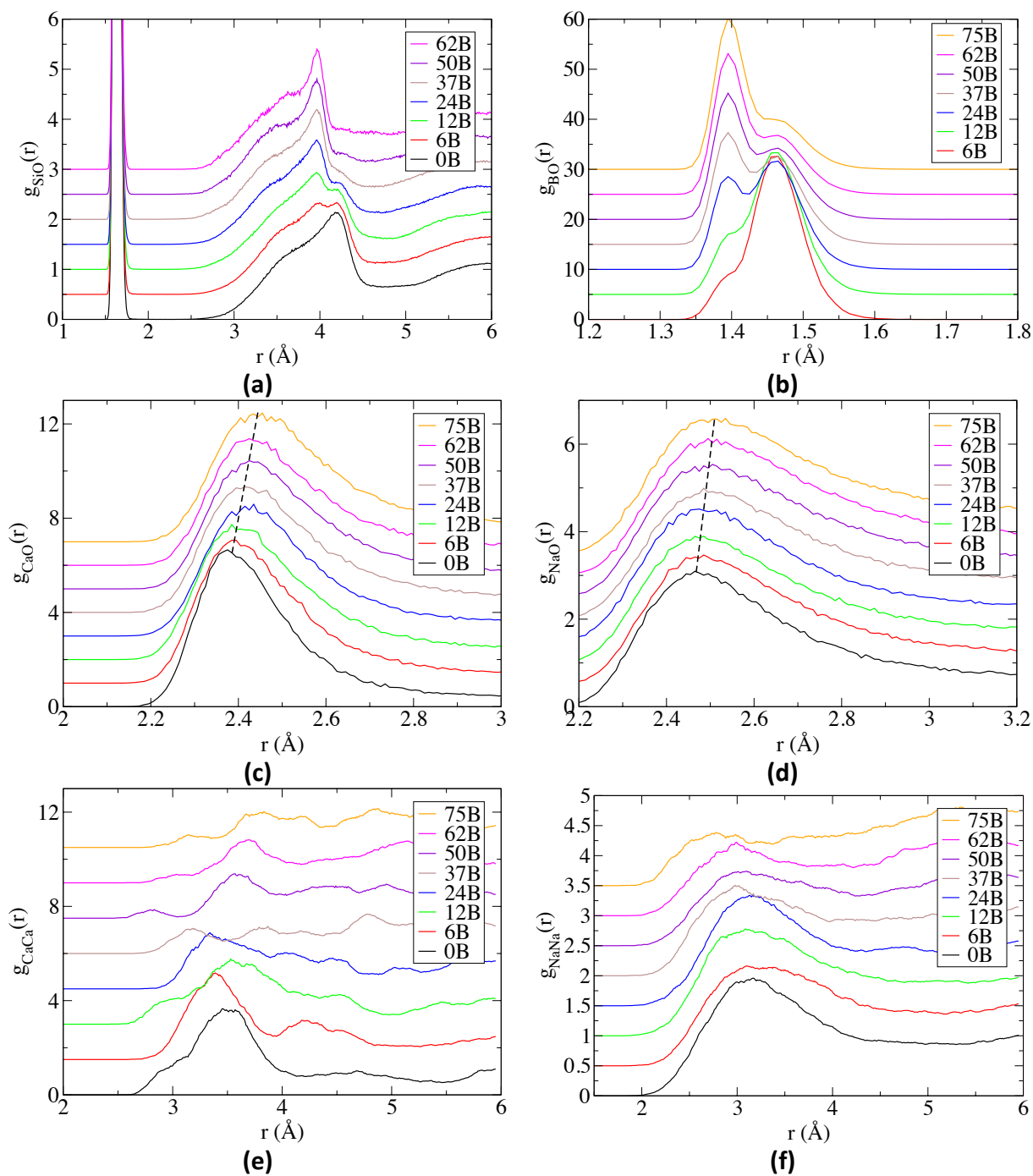

**Fig. 4:** **(a)** Si–O, **(b)** B–O, **(c)** Ca–O, **(d)** Na–O, **(e)** CaCa, and **(f)** Na–Na partial pair distribution functions of borosilicate glasses with varying B/Si molar ratios (see **Tab. 1**).

### 3.3 Bond angle distributions

We now investigate the angular environment of each element. As in the case of the B–O PDF, the O–B–O bond angle distribution exhibits a bimodal distribution (see **Fig. 5a**). This arises from the gradual increase in the population of 3-fold coordinated B atoms at the expense of 4-fold coordinated species as the amount of $B_2O_3$ increases (see **Fig. 1b**). We observe that the average O–B–O bond angle is around 109° and 120° for 4- and 3-fold coordinated B atoms, as expected from their tetrahedral and trigonal geometry, respectively. As shown in **Fig. 5b**, the angular environment of 3- and 4-fold coordinated B atoms remains largely unaffected by composition. A similar situation is observed for the O–Si–O intra-tetrahedral angle, which remains constant for all the simulated glasses and is found to be around 109°, in agreement with the tetrahedral geometry of $SiO_4$ units (see **Fig. 5c**).

In contrast, the inter-polytope X–O–X angle (where X = Si or B) formed by BO atoms exhibits some significant variations with composition (see **Fig. 5d**). As shown in **Fig. 6**, the average value of the inter-polytope angle formed around BO atoms tend to initially decrease upon the increase in $SiO_2$ concentration, reaches a minimum at around $[SiO_2]/([SiO_2]+[B_2O_3]) = 0.7$, and eventually tend to increase again. Interestingly, this trend mirrors that of the density, namely, the maximum of density corresponds to the minimum of the BO angle (see **Fig. 1a**). This suggests that the angle formed around BO atoms has a strong influence on the density, that is, lower BO angle values result in larger density values. This can be understood from the fact that lower BO angles enable an enhanced compactness of the atomic network, that is, with less empty space between the $SiO_4$, $BO_3$, or $BO_4$ polytopes. Interestingly, the glass hardness was also observed to be maximum in the range of compositions wherein the BO angle reaches a minimum [26], which, as suggested before, confirms the important role of the inter-tetrahedral angle in controlling the mechanical properties of silicate glasses [48,49].

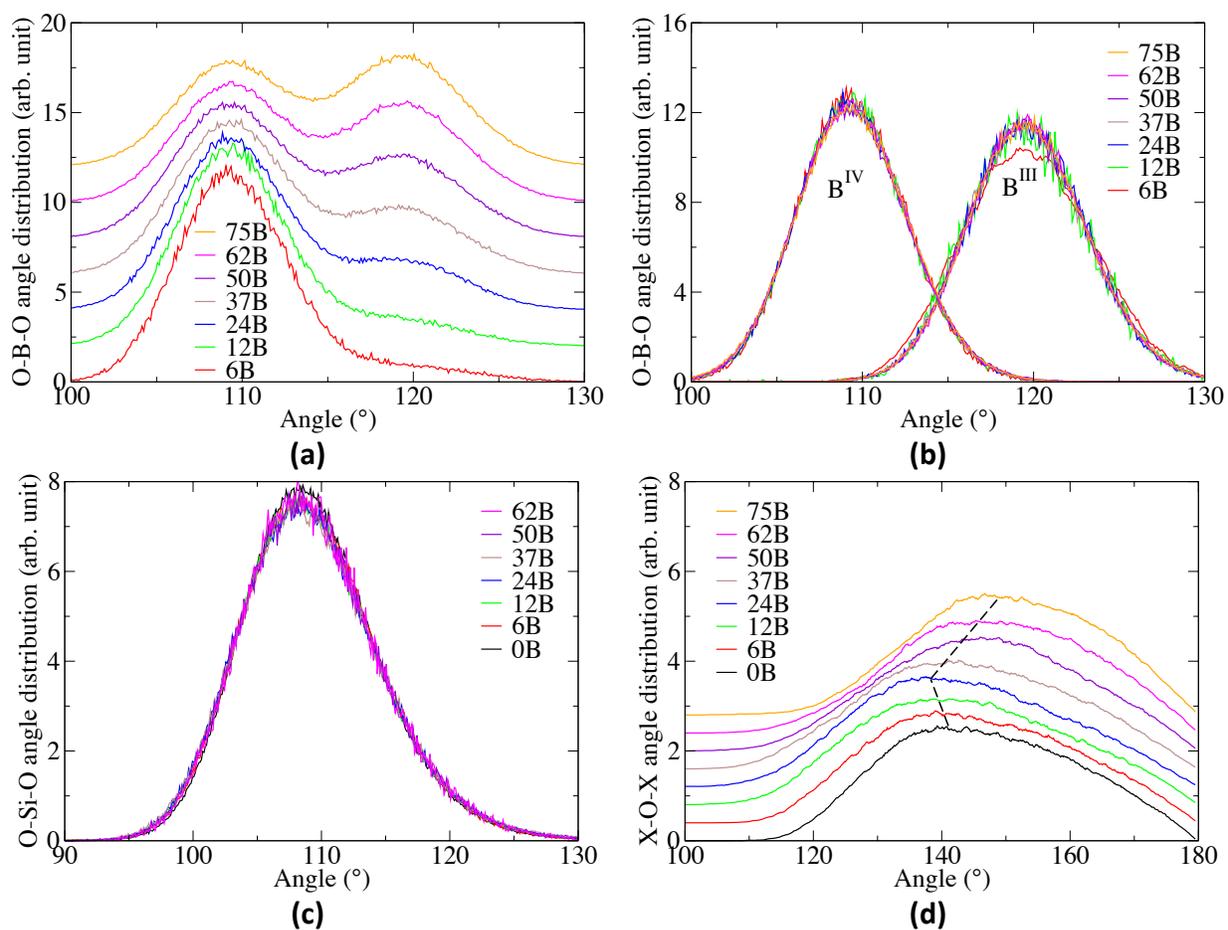

**Fig. 5:** Computed **(a)** O–B–O, (b) O–B$^{III}$–O, O–B$^{IV}$–O, (c) O–Si–O, and (d) X–O–X (where X = Si or B) bond angle distributions of borosilicate glasses with varying B/Si molar ratios (see **Tab. 1**). The dashed line is a guide for the eye.

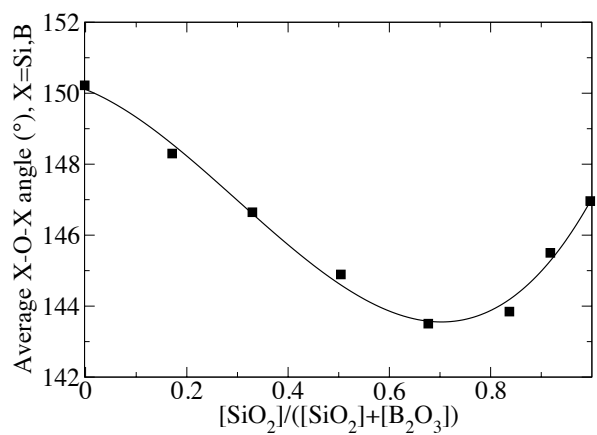

**Fig. 6:** Average X–O–X (where X = Si or B) angle as a function of composition. The line is a guide for the eye.

## 3.4 Structure factor

Although it contains the same information as the pair distribution function, the structure factor places more emphasis on large distances and, hence, can be used to investigate the medium-range order structure of borosilicate glasses. **Fig. 7** shows the neutron structure factor (see **Sec. 2.3**) computed for the 10B glass (see **Tab. 1**) compared with experimental neutron diffraction data [13]. We observe a very good agreement between simulated and experimental data. The intensity and position of the diffraction peaks are well predicted, both at low and high $Q$ values. Specifically, we note that the first sharp diffraction peak of the structure factor is very well reproduced, which suggests that the medium-range order structure of the glass is well reproduced. For comparison, we note that Kieu's potential yields roughly the same level of agreement with experimental data.

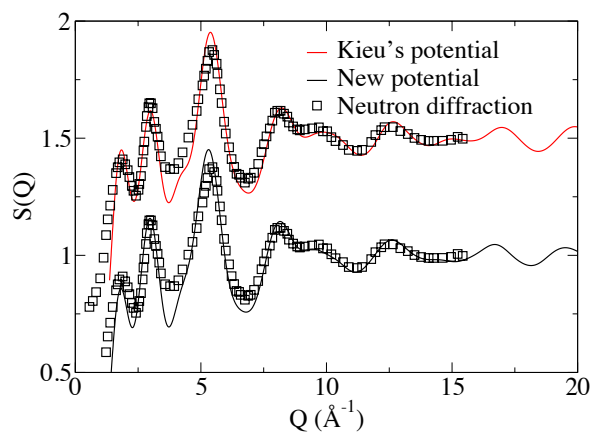

**Fig. 7:** Total neutron structure factor of the 10B glass computed with our new potential and compared with experimental neutron diffraction data [13]. The total neutron structure factor of the same glass computed with Kieu's potential [23] is also presented and compared with the same experimental data (vertically shifted by +0.5).

## 3.5 Ring size distribution

We further investigate the medium-range order structure of the borosilicate glasses by computing the ring size distribution (see **Sec. 2.3**). **Fig. 8a** shows the computed ring size distributions, which are, overall, in agreement with previous simulation studies [24,25,50,51]. Starting from the silicate glass (0B), we observe that the insertion of $B_2O_3$ first results in the creation of additional short rings (with a size lower than 6). However, as the amount of $B_2O_3$ keeps increasing, we eventually observe the formation of large rings (with a size larger than 10) that are nearly absent in the silicate glass. Altogether, these effects results in a minimum in the average ring size around $[SiO_2]/([SiO_2]+[B_2O_3]) = 0.7$ (see **Fig. 8b**).

This can be understood from the fact that, at low amount of $B_2O_3$, B atoms take the form of $BO_4$ units that are charge compensated by Na or Ca cations (see **Fig. 1b**). The addition of each B atom thereby effectively consumes one NBO, as Na or Ca turn from a network-modifying to a charge-compensating role. This induces an increase in the overall connectivity of the glass, which results in a decrease in the average ring size. In contrast, with higher amount of $B_2O_3$, B atoms start to take the form of $BO_3$ units (see **Fig. 1b**). At this point, the addition of B atoms does not consume NBOs any longer and effectively decreases the overall connectivity of the network, as 4-fold coordinated Si are replaced by 3-fold coordinated B atoms. This results in an increase in the average ring size.

We note that the composition (24B) at which the average ring size is minimum also exhibits a maximum in density and a minimum in the average BO inter-polytope angle. This arises from the fact that smaller rings result in a more compact (denser) network. This also suggests a close connection between the size of the rings and the average value of the average BO inter-polytope angle.

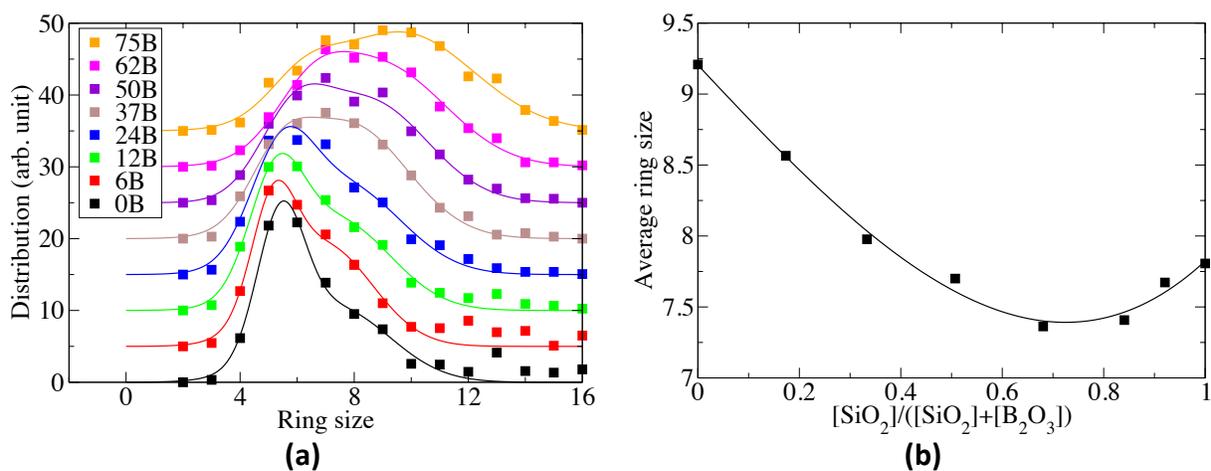

**Fig. 8: (a)** Computed ring size distributions of borosilicate glasses with varying B/Si molar ratios (see **Tab. 1**). **(b)** Average ring size as a function of composition. The lines are guides for the eye.

### 3.6 Shear viscosity

The shear viscosity is a property that is very sensitive to the details of the empirical potential and, hence, is traditionally challenging to predict using classical MD [40]. To assess the ability of our new potential to predict realistic dynamic properties, we compute the isothermal shear viscosity of the modified borosilicate systems selected for this study (see **Sec. 2.3**). **Fig. 9** shows the shear viscosity computed at 1150 °C. Overall, we observe that the viscosity tends to increase with the concentration of $SiO_2$.

To validate this trend, we estimate the viscosity of the borosilicate glass-forming systems using the MYEGA equation of equilibrium viscosity [52] by taking the experimental glass transition temperature and fragility index of the borosilicate glasses as inputs [26]. The comparison between experimental (i.e., based on the MYEGA equation) and computed results is presented in **Fig. 9**. We observe that our new potential systematically underestimates the experimental shear viscosity. This appears to be a general limitation of classical empirical interatomic potential, which often overestimates the fluidity of supercooled liquids [40,47,53]. However, the overall trend of the computed viscosity data is in good agreement with that of the experimental data, which suggests that our new potential yield a reasonable description of the dynamics of the system.

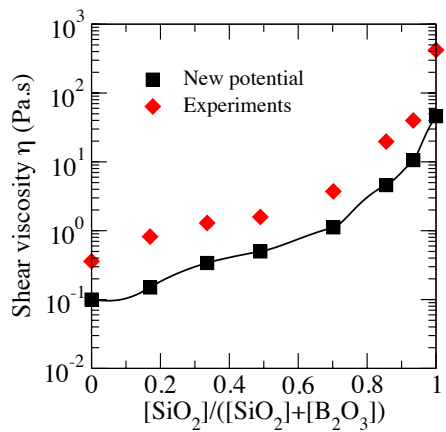

**Fig. 9:** Computed shear viscosity (at 1150 °C) of borosilicate glass-forming systems with varying B/Si molar ratios (see **Tab. 1**) compared to experimental data [26]. The line is a guide for the eye.

# 4. Discussion

## 4.1 Performance of the new potential

Overall, we observe that our new potential yields a good agreement with experiments in terms of: (i) overall short-range order, (ii) overall medium range order, (iii) B atom coordination, (iv) glass density, and (v) supercooled liquid viscosity. This demonstrates that the use of composition-dependent partial charges and potential parameters is not absolutely required to develop a reliable potential to model borosilicate glasses.

Although the excellent agreement between computed and experimental data regarding the coordination number of B atoms suggests that our new potential properly describes B–O interactions, it must be noted that this agreement might be partially artificial. Indeed, by construction, our potential is artificially "forced" to reproduce the evolution of the coordination number of B atoms with composition. However, this quantity is known to be sensitive to the thermal history of the glass [21,26]—hyperquenched glasses that would be generated with the cooling rate used herein (1 K/ps) are expected to exhibit lower fractions of 4-fold coordinated B atoms. Hence, it is likely that this apparent agreement between experimental data and the outcomes of our new potential might arise from a "cancellation of errors." However, since classical molecular dynamics is intrinsically limited to the use of high cooling rate due to computational constraints, such mutual cancellations of errors may be necessary to generate a glass structure that can meaningfully be compared with experimental data obtained on glasses generated with slower cooling rates.

The use of partial charges is worth some discussion. It is now well-established that empirical potential relying on partial charges, rather than the formal charges (i.e., −2 for O atoms), yield better results [14]. This is due to their ability to partially capture the complex delocalization of the electronic density in ionocovalent bonds, which occurs via charge transfer and polarization effects [14]. In contrast, potentials relying on formal charges usually require the use of 3-body interaction terms to properly describe the structure of silicate glasses [20,54]. It is interesting to note that our new potential relies on a partial charge for O atoms of –0.945, which is reminiscent of the original GS potential [39]. This value was also used by Matsui *et al.* to develop a transferable interatomic potential for CaO–MgO–Al$_2$O$_3$–SiO$_2$ crystals and melts [55]. This suggests that this specific set of partial charges shows a great transferability over modified silicate and borate systems. In contrast, the well-known BKS potential relies on a partial charge of –1.2 for O atoms and describes well the structure of glassy silica [56–58]. However, a recent reparametrization of the BKS potential based on the pair distribution obtained by *ab initio* simulations led to a partial charge of –0.9552 [59], which is very close to the value used herein. Hence, the use of a partial charge that is smaller (in absolute value) than –1.2 for O atoms, which yields softer interatomic energies, appears to offer a more realistic description of silicate melts and glasses. This might arise from an effect of temperature on charge transfer and polarization effects, which is usually ignored when potentials are fitted against crystal properties at zero temperature.

## *4.2 Relevance for nuclear waste immobilization glasses*

We now discuss the importance of developing accurate interatomic potentials to investigate the structure and properties of nuclear waste immobilization glasses, which is a technology important application of borosilicate glasses. Vitrification is usually considered as the method of choice to immobilize high-level nuclear waste due to: (i) the high capacity of glasses to accommodate and immobilize a large range of radioactive elements, (ii) the small volume of the resulting wasteforms, (iii) the high tolerance of glass to radiation damage, and (iv) the high chemical durability of glasses in aqueous conditions [9]. To the end, borosilicate glasses are commonly used thanks to their excellent properties [60].

The development of novel borosilicate glasses featuring improved performance for nuclear waste immobilization has to result from a compromise between the ability to achieve high loadings of wastes, technical constraints linked to the processability of the glass, and glass performances. As such, optimal glasses must simultaneously balance various constraints: (i) to allow high waste loading, so that waste glass volumes and the associated storage and disposal costs are reduced, (ii) to offer an easy processing (vitrification should be performed at relatively low temperatures because of the volatility of the fission products), (iii) to exhibit a controlled viscosity (the viscosity should be controlled to ensure high throughput and controlled pouring at the processing temperature), (iv) to remain homogeneous and avoid uncontrolled crystallization or phase separation, (v) to feature high stability with respect to radiations, temperature, or mechanical stress, (vi) to comply with regulations, and (vii) to show an excellent chemical durability (which is of primary importance to ensure low release rates for radionuclides on potential contact with water).

For all the reasons previously discussed, it is important to understand how glass composition control its engineering properties (e.g., glass transition temperature, viscosity, propensity for crystallization or phase separation, stiffness, toughness, thermal expansion, or dissolution rate).

However, the knowledge of the disordered glass structure often acts as a missing link between composition and properties. To this end, the development of accurate empirical potential may be key, as most conventional experimental techniques do not provide a direct and full access to the atomic structure of borosilicate glasses.

In particular, a topological model of dissolution has recently been developed to predict the dissolution rate of modified oxide glasses and has met some great success [61–66]. However, this model requires an accurate knowledge of the glass connectivity, including the average coordination number of B atoms. Classical molecular dynamics can be key to elucidate the structural inputs that are needed to parametrize such models.

In addition, among the different properties of supercooled liquids, the shear viscosity is of special interest, as it controls the processability of the melt. Indeed, all the stages of the glass production (melting, mixing, and fining to the final forming) require a careful control of the shear viscosity. In the context of nuclear waste immobilization glasses, current waste vitrification technologies typically operate at temperatures between 1150 and 1250 °C [67]. Such temperature must be kept low to avoid the loss of volatile components and save costs, but should be high enough to ensure small values of viscosity. In practice, the viscosity at the processing temperature is typically maintained: (i) above 2 Pa·s, as glasses with a lower viscosity tend to penetrate into the melter bricks and to excessively corrode melter walls and (ii) under 11 Pa·s to ensure high throughput and controlled pouring into canisters [67]. Indeed, to minimize operational costs, an efficient vitrification process requires high throughput, which increases with decreasing viscosity at the processing temperature. For all these reasons, it is important to understand the link between glass composition and shear viscosity of supercooled liquids. For instance, in the case of the glasses simulated herein, only two glasses out of eight (6B and 12B) would exhibit acceptable viscosity values. This illustrates how the use of reliable molecular dynamics simulations could be key to quickly screen promising glasses exhibiting desirable viscosity values.

## 5. Conclusion

We report the development of a new empirical interatomic potential for borosilicate glasses. Although relying on a simple formulation and constant parameters, our new potential offers a very good agreement with experimental data of glass structure and properties. In particular, our potential yields realistic average B coordination number, glass density, overall short-range and medium-range order structure, and shear viscosity values for a large number of borosilicate glasses and liquids. This denotes the good transferability of our potential over a range wide of glasses, from silicates to borates.

## Acknowledgements

MB gratefully acknowledges invaluable discussions with B. Guillot, N. Sator, and M. Micoulaut over the years regarding the performances and limitations of empirical potentials for molecular dynamics. This work was partially funded by Corning Incorporated and National Science

Foundation under Grant No. 1562066. M.M.S. acknowledges support from the Independent Research Fund Denmark (Grant No. 7017-00019).

Foundation under Grant No. 1562066. M.M.S. acknowledges support from the Independent Research Fund Denmark (Grant No. 7017-00019).